\documentclass[aps,prd]{revtex4}  
\usepackage[latin1]{inputenc}  
\usepackage{hyperref}  
\usepackage{amsmath}  
\usepackage{graphicx}  
\usepackage{amssymb}

\date{\today}  
\def\be{\begin{equation}}  
\def\bea{\begin{eqnarray}}  
\def\eea{\end{eqnarray}}  
\def\ee{\end{equation}}  
\def\no{\nonumber}  
\def \a{\alpha}  
\def \O{\Omega}  
\def \d{\delta}  
  
\begin{document}  
\title{Reducing the flexing of the arms of LISA}  
\author{K. Rajesh Nayak$^{1}$, S. Koshti$^2$, S. V. Dhurandhar$^2$ and J-Y. Vinet$^3$}  
\affiliation{ $^1$The University of Texas at Brownsville,   
\\80 Fort Brown, Brownsville, TX 78520, USA.  
\\ $^2$IUCAA, Postbag 4, Ganeshkind, Pune - 411 007, India.   
\\ $^3$ARTEMIS, Observatoire de la Cote d'Azur,   
\\ BP 4229, 06304 Nice, France.  
}  
  
\begin{abstract}  
The joint NASA-ESA mission LISA relies crucially on the stability of the three   
spacecraft  constellation. All three spacecraft are on heliocentric and weakly eccentric orbits   
forming a stable triangle.  It has been shown that for certain spacecraft orbits,   
the arms keep constant distances to the first order in the eccentricities.   
However, exact orbitography shows the so-called `breathing modes' of the arms where the arms   
slowly change their lengths over the time-scale of a year. In this paper we   
analyse the breathing modes (the flexing of the arms) with the help of the   
geodesic deviation equations to octupole order which are shown to be equivalent   
to higher order Clohessy-Wiltshire equations. We show that the flexing   
of the arms of LISA as given by the `exact' solution of Keplerian orbits,   
which gives constant armlengths to the first order in eccentricity and whose maximum flexing   
amplitude is $\sim 115,000$ km, can be improved, by tilting the plane   
of the LISA triangle slightly from the proposed orientation of $60^\circ$ with   
the ecliptic to obtain a maximum flexing amplitude of $\sim 48,000$ km, reducing it  
by a factor of $\sim 2.4$. The reduction factor is even larger if we consider  
the corresponding Doppler shifts, for which the reduction factor reaches almost  
a factor of 6. We solve the second order equations and obtain the general   
solution. We then use the general solution to establish the optimality of the solutions that we have found.     
\end{abstract}  
  
\maketitle  
\section{Introduction \label{SC:1}}  
  
Detection of low frequency (below 100 mHz) gravitational waves (LFGW) is a challenging endeavour   
because  
terrestrial instruments are dominated by seismic noise in this frequency range. There is not much    
hope of isolating the noise in this frequency range, which mainly arises from the direct Newtonian   
coupling of ground motions with the test masses. This is why space missions have been given serious   
considerations from early on, because LFGW are extremely important in the physics of and around   
massive black holes which are hosted by most of galaxies, including ours.   
The LISA mission \cite{RIP} involves three spacecraft forming a triangle of side   
$\sim 5 \times 10^6$ km, linked by optical beams. The gravitational wave (GW) signal is   
read from the beat note generated between an incoming beam and the local laser, so that six such   
signals are generated between the three pairs of spacecraft.   
In a classical Michelson  
interferometer, the requirements on the laser phase noise are strongly reduced by the symmetry   
between the two arms. In the case of LISA, a comparison of the laser beam from a distant spacecraft   
with the local oscillator would demand a relative frequency stability better than the GW amplitude,   
which is quite unrealistic. This was realised from the very beginning. The idea which allows   
relaxing the requirement of frequency stability is to digitally reproduce an interferometric   
configuration by mixing the six elementary data flows with delays corresponding to virtual   
optical paths between spacecrafts. The corresponding technique is called Time-Delay Interferometry   
(TDI) \cite{TDI}. But the practical realisation of a triangular configuration orbiting   
the sun leads to a structure only approximately rigid. In fact, the three spacecraft are on   
almost circular orbits of small eccentricity $e$, and the mutual distances are constant during   
the year only to the first order in $e$. When an exact calculation of the mutual distances between   
pairs of spacecraft is performed using Keplerian orbital equations, a `breathing' motion of the   
arms appears, normally termed as `flexing' of the arms, with amplitude of  about 115,000 km.   
This relative motion between spacecraft causes a Doppler effect, but at   
frequency ($\sim 10^{-7}$ Hz) out of the detection band $10^{-4} - 10^{-1}$ Hz.    
It has nevertheless been shown \cite{TEA04} that this motion    
makes the delay operators of TDI, variable in time and thus prevents the exact cancellation of the   
laser frequency noise.  
One can however, suppress the noise but at the cost of complex application   
of TDI techniques involving noncommuting delay operators. When first generation TDI  
combinations are used, the residual noise is proportional to the time derivative  
of the flexing, so that a significant reduction of the slow orbital Doppler effect would make 
it necessary to reconsider the TDI strategy. We have found it useful, therefore, to   
discuss the stability of the LISA triangle to higher orders in $e$ instead of only to the first   
order, in order to investigate whether better orbits are possible, with reduced breathing amplitude. 
\par 
The paper is organised as follows. In section \ref{SC:2}, we firstly show that a small correction to the angle between the LISA plane and the ecliptic is able to reduce by a factor $\sim 2.4$ the RMS flexing amplitude and more importantly the peak to peak Doppler shift by a factor $\sim 5.6$ which is relevant for TDI. In section \ref{SC:3} we show how a generalisation of the Clohessy-Wiltshire equations up to second order, namely, the octupolar   
expansion of the Newtonian potential and the related geodesic deviation, allows us to reproduce   
almost exactly the behaviour as obtained from the exact orbits. This paves for us a natural way   
for introducing extra degrees of freedom in the orbital model in order to establish the optimality   
of the solution we have found. In section \ref{SC:4}, we achieve this by obtaining the general solution to the   
second order equations and vary the arbitrary constants to arrive at the optimal results.    
In this analysis, for simplicity, we have ignored the effects of the gravitational pulls of   
Jupiter and Earth which are of the same order. We note that the flexing occurs mainly due to two reasons: (i) the non-existence of exact Keplerian orbits of spacecraft allowing constant arm-lengths, and (ii) the perturbation due to planets in our Solar system. In this paper we address the first issue only.  
   
\section{Reducing the flexing of the `exact' solution by slightly tilting the plane \label{SC:2}}  
  
The LISA mission requires that the distances between the three spacecraft   
remain constant to the first order in the parameter $\a = l /2R$. $\a$ is proportional to the   
eccentricity $e$ to the first order: $\a = \sqrt{3} e$ to this order in $e$.   
We take $l = 5 \times 10^6$ km and $R = 1 $ A. U. $\sim 1.5 \times 10^8$ km, and thus   
$\a \sim 1/60$.   
The constancy of armlengths alone does not lead to a unique choice of the orbits.   
It has been shown that all the points around a circular  reference orbit,   
in a plane making  an angle of $\pm \frac{\pi}{3}$ with respect to the   
plane of  reference orbit satisfies the above condition\cite{DNKV,PB,FHSVB,H,HF,THS}.  
However, all these orbits have different arm length variation to the  
second order in $\alpha$, also known as flexing.  
In reference \cite{DNKV} (henceforth referred to as paper I),   
we have presented a set  of elliptical orbits satisfying the above requirement.   
Here  we extend our analysis to second order in $\alpha$ which describes the flexing of the   
arms quite accurately. We show this by comparing the second order results with the exact results   
which agree to within few parts in $10^4$. Thus the second order analysis allows us to    
handle the flexing in a convenient and reliable manner. The eventual aim of this excercise   
is to minimise the flexing which could be important for the exploitation of LISA.  
\par  
For this purpose we need to decide on the measure of flexing. There are several ways of   
characterising this quantity. We can take this to be the peak to peak variation of the   
armlength - the difference between the maximum distance between two given spacecraft and the minimum   
distance between the two spacecraft over the period of one year; or we can define   
it as a r.m.s variation of armlength over the period of a year. An important measure is the Doppler shift which depends on the relative velocity between two spacecraft. We investigate 
all these measures of flexing in this paper.  
We also make the simplifying assumption that the orbits of the spacecrafts 2 and 3 are   
obtained by rigidly rotating the orbit of spacecraft 1 by $120^\circ$ and $240^\circ$   
(this is described more precisely in the next subsection). Due to the symmetry assumed   
in this model, the measures of flexing although different from each other are   
independent of the spacecraft pair we choose. Therefore without loss of generality we consider    
spacecraft 1 and spacecraft 2. If $l_{12}(t)$ is the distance between the spacecraft   
1 and 2 at time $t$, then the peak to peak flexing is equal to   
$\max{l_{12} (t)} - \min{l_{12} (t)}$, where the maxima and minima are taken over the period   
of one year. The $l_{23} (t)$ and $l_{13} (t)$ are time-shifted versions of $l_{12} (t)$   
by one third and two thirds of an year respectively. The goal of this section is to explore   
the possibility of reducing the flexing by tweaking the orbital parameters of the spacecraft   
orbits. One way is to vary the tilt of the plane of LISA and study the effect on the arm   
length variation as a function of the tilt angle and then minimise the flexing by appropriately   
choosing the tilt. We show that the flexing can be reduced by appropriately choosing the tilt   
and the flexing can be reduced to a minimum.  The generic problem involves exploring the space   
of orbital parameters of the three spacecraft orbits. Because of the symmetry assumed here,   
we are restricting ourselves to a subspace, whose dimensionality is smaller by a factor of   
three than the dimensionality of the full parameter space of orbits. Following  paper I,   
we start with the exact orbital equations.  
\subsection{The exact orbits \label{SC:2A}}  
\begin{figure}[t]  
\begin{center}  
\includegraphics[width=0.8\textwidth]{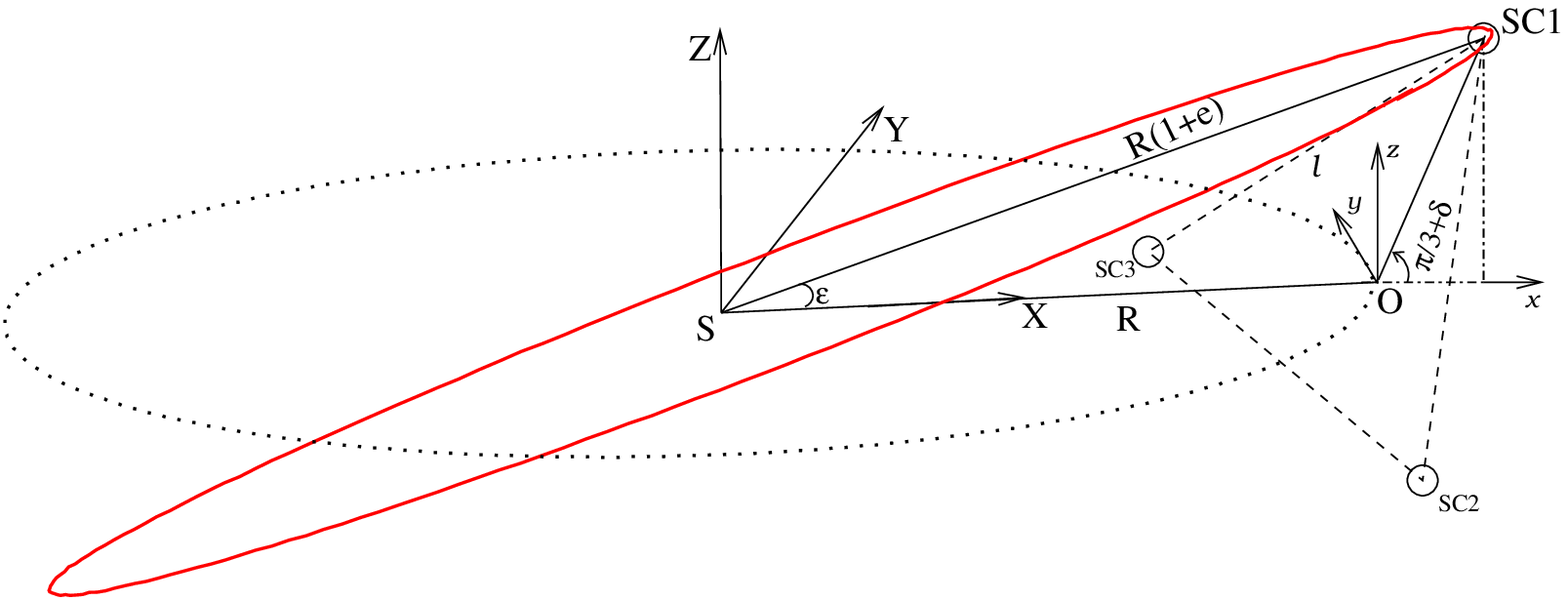}  
\caption{The figure shows the geometry of the orbits and of LISA.   
The barycentric frame is labelled by $\{X, Y, Z\}$ while the CW frame   
is labelled by $\{x, y, z\}$. SC1, SC2 and SC3 denote the three spacecraft.   
The radius of the reference orbit is taken to be $R = 1$  A. U. and S   
denotes the Sun.}  
\label{geom}  
\end{center}  
\end{figure}

We choose the barycentric frame with coordinates $\{X, Y, Z\}$ as follows:   
The ecliptic plane is the $X-Y$ plane and we consider a circular reference orbit   
of radius 1 A. U. centred at the Sun.  Let $\delta $ be a small correction of the order   
$\alpha$, to the tilt of the LISA plane.   
We choose the axes so that the tilt is $\pi/3 + \delta$.  
We choose spacecraft 1 to be at its highest point (maximum Z) at $t = 0$.   
This means that at this point, $t = 0$ and $Y = 0$. The   
orbit of the first spacecraft is an ellipse with inclination angle   
$\epsilon$,  eccentricity $e$ and satisfying the above initial condition.  
The geometry of the configuration is shown in  
Figure  \ref{geom}.    
From the geometry, $\epsilon$ and $e$ are obtained as functions of $\delta$,  
\bea  
\tan \epsilon &=& \frac{2}{\sqrt{3}} \frac{\alpha \sin(\frac{\pi}{3}+\delta)}  
{\left[ 1+\frac{2}{\sqrt{3}} \alpha \cos(\frac{\pi}{3}+ \delta )\right]} \, ,\nonumber  \\   
e &=& \left[ 1+ \frac{4}{3} \a^2   + \frac{4}{\sqrt{3}}\a   
\cos \left( \frac{\pi}{3}+\delta \right) \right]^{1/2} - 1 \, ,  
\label{eq:eincl}  
\eea  
and the orbit equations for the spacecraft 1 are given by:  
\bea  
X_1 &=& R(\cos \psi_1 + e) \cos \epsilon, \no \\  
Y_1 &=& R \sqrt{1 - e^2} \sin \psi_1, \no \\  
Z_1 &=& R(\cos \psi_1 + e) \sin \epsilon.   
\label{tltorb}  
\eea    
The eccentric anomaly $\psi_1$ is implicitly given in terms of $t$ by,  
\be  
\psi_1 + e \sin \psi_1 = \Omega t \, ,  
\ee  
where $t$ is the time and $\Omega$ is the average angular velocity.   
The orbits of the spacecraft 2 and 3 are obtained by rotating the orbit   
of spacecraft 1 by   
$2 \pi / 3$ and $4 \pi/3$ about the $Z-$axis; the phases $\psi_2, \psi_3$,    
however, must be adjusted so that the spacecraft are about the distance $l$   
from each other. The orbital equations of spacecraft $k = 2, 3$ are:  
\bea  
X_k &=&  X_1 \cos \sigma_k- Y_1 \sin \sigma_k \, , \no \\  
Y_k &=&  X_1 \sin \sigma_k +  Y_1 \cos \sigma_k \, , \no \\  
Z_k &=& Z_1 \, ,  
\label{orbits}  
\eea  
where $\sigma_k = \left(k-1\right) \frac{2 \pi}{3}$,  
with the caveat that the $\psi_1$ is replaced by the phases $\psi_k$, where they   
are implicitly given by,  
\be  
\psi_k + e \sin \psi_k = \Omega t - (k -1) \frac{2 \pi}{3}= \Omega t-\sigma_k.  
\label{eq:psik}  
\ee   
These are the exact orbital equations for  the three spacecraft.  
\par   
We investigate the arm length variation as function of the tilt angle $\delta$.  
The distance between spacecraft 1 and 2,    
$l_{12} = \{{(X_1 - X_2)^2 + (Y_1 - Y_2)^2 + (Z_1 - Z_2)^2}\}^{1/2}$,   
as a function of $t$ and $\d$ is shown in Figure 2 (a). Figure 2 (b) shows constant   
$\delta$ sections of the 3 D plot in (a). For $\d = 0$ the peak to peak variation in   
armlength is about 115,000 km while the r.m.s. variation is about 36000 km. However,   
as can also be seen from the plots, the variation in armlength can be reduced by a   
factor little more than 2. Our goal is to determine the tilt angle which minimises   
the flexing and also to determine this minimum. Our second goal is to show that   
sufficiently accurate results can be obtained by going only to the second order in $\a$.   
We start from the exact orbits and then make a second order expansion in $\a$ in the   
next subsection.  
\begin{figure}[t]  
\begin{center}  
\includegraphics[width=1\textwidth]{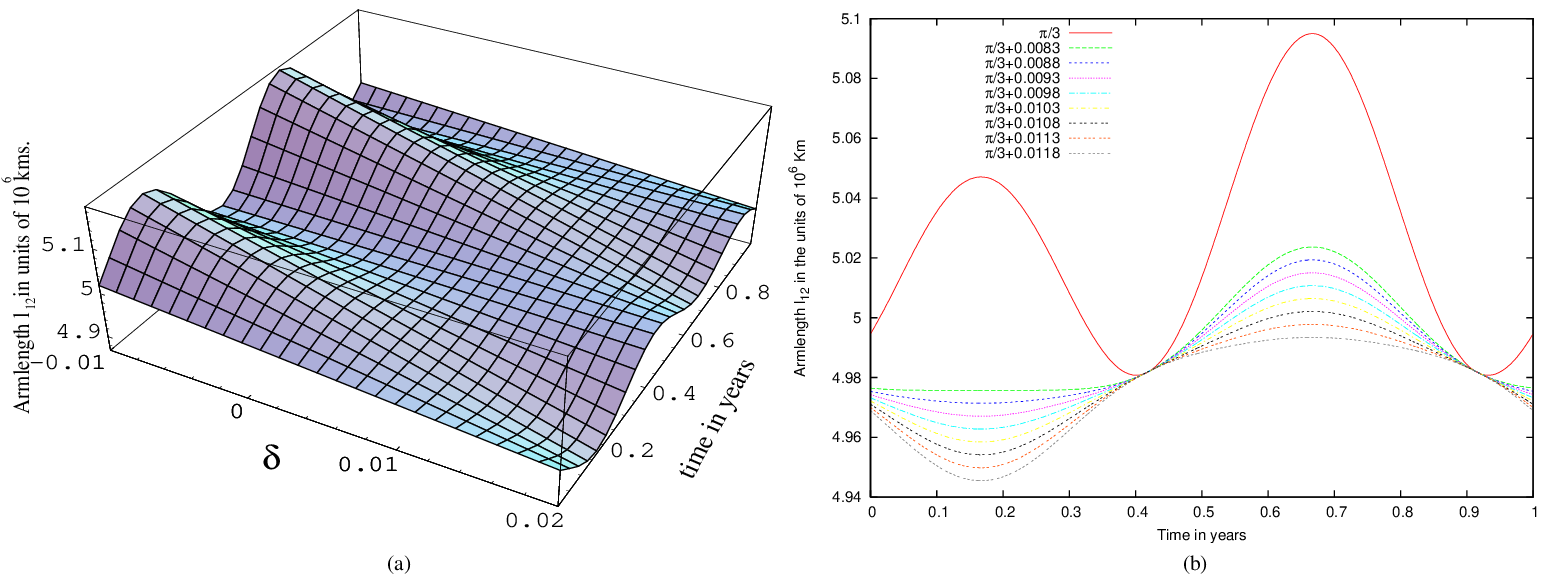}  
\caption{  
(a) \  
Plot of distance $l_{12}$ as a function of time and tilt angle  
$\delta$.  
(b) \ Plot of the optimal variation in arm length as function of time.  
	For comparison we have also given $l_{12}$ for tilt equal to $\frac{\pi}{3}$.  
}  
\label{l12td}  
\end{center}  
\end{figure}    
\subsection{The orbits to second order in $\alpha$ \label{SC:2B}}  
In this section, we obtain the expression for orbits to the second order in   
$\alpha$. From the exact expressions for the inclination  $\epsilon$   
and the eccentricity $e$ given in Eq. (\ref{eq:eincl}), to second order  
in $\a$ we obtain,  
\begin{eqnarray}  
	e & = & \frac{\alpha}{\sqrt{3}} + \frac{\alpha^2}{2}-\alpha\delta\, , \nonumber\\  
	\epsilon & = & \alpha - \frac{\alpha^2}{\sqrt{3}}+\frac{\alpha \delta}{\sqrt{3}}\,  
.\label{eq:epsl}  
\end{eqnarray}  
Eq. (\ref{eq:psik}) can be solved iteratively to the required order in $e$ to   
obtain an expression for  $\psi_k(t)$ as a function of time, substituting for   
$e$ from Eq. (\ref{eq:epsl}) and retaining terms only to the second order in $\alpha$.   
We then have,   
\begin{equation}  
	\psi_k(t)=\phi_k-\frac{\alpha}{\sqrt{3}} \, \sin\phi_k \, +\,  
	\frac{\alpha^2}{6} \,  \left(  \sin2\phi_k\, -\,3   
	\sin\phi_k \right)+\alpha\delta\, \sin\phi_k,  
\end{equation}  
where $\phi_k=\Omega t- \frac{2\pi}{3}(k-1)=\Omega t - \sigma_k $, for the   
$k$th spacecraft. The orbital equation for the $k$th spacecraft can be   
obtained by substituting these values in Eq.(\ref{orbits}). The expressions take on   
simpler form if we transform to frames tied to LISA. Here we use these approximate,   
albeit, simpler equations in order to compute the inter-spacecraft distances.   
The first transformation is to the Clohessy-Wiltshire (CW) frame.  
\par  
Clohessy and Wiltshire \cite{CW} make a transformation to a frame - the CW frame $\{x, y, z\}$   
which has its origin on the reference orbit and also rotates with angular velocity $\Omega$.    
The $x$ direction is normal and coplanar with the reference orbit, the $y$ direction is  
tangential and comoving, and the $z$ direction is chosen orthogonal to the orbital plane.    
They write down the linearised dynamical equations for test-particles in the   
neighbourhood of a reference particle (such as the Earth). Since the  
frame is noninertial, Coriolis and centrifugal forces appear in addition to the tidal forces.   
\par
We take the reference particle to be orbiting in a circle of radius $R$ with constant angular velocity 
$\Omega$. Then the transformation to the  CW frame $\{ x, y, z \}$ from the barycentric frame $\{ X, Y, Z \}$ is given by,  
\begin{eqnarray}  
x & = & \left(X-R\,\cos\Omega t\right)\,\cos\Omega t\;+\;\left(Y-R\,\sin\Omega t\right)\,  
\sin\Omega t\,,\nonumber \\  
y & = & -\left(X-R\,\cos\Omega t\right)\,\sin\Omega t\;+\;\left(Y-R\,\sin\Omega t\right)\,  
\cos\Omega t\,,\nonumber \\  
z & = & Z.\label{eq:CW}  
\end{eqnarray}  
The  second transformation to the LISA frame is given by rotating the CW frame about   
$y$ axis by an angle $\pi/3 + \delta$. With these two transformations the orbital  
equations for the $k$th spacecraft to the second order in $\a$ are given by,  
\begin{eqnarray}  
x'_{k} & = & \frac{2}{\sqrt{3}}R\,\alpha\,\cos\phi_{k}\,+\,\frac{1}{24}R\,\alpha^{2}  
\left[13 - 6 \cos\phi_k - 7 \cos2\phi_k \right]\,,\nonumber \\  
y'_{k} & = & -\frac{2}{\sqrt{3}}R\,\alpha\,\sin\phi_{k}\:+  
2 R \alpha \delta \sin\phi_k \, +\,    
\frac{1}{3}R\,  
\alpha^{2}\,\left[\sin2\phi_{k}\:-\:3\,\sin\phi_{k}\,\right]\,,\nonumber \\  
z'_{k} & = & \frac{1}{8\sqrt{3}}R\,\alpha^{2}\,\left[11\,-\,  
10\,\cos\phi_{k}\,-  
\cos2\phi_{k}\right]\,.\label{eq:xpk}  
\end{eqnarray}  
Using these equations it is easier to compute the distance between spacecraft.   
  
\subsection{Optimal tilt angle \label{SC:2C}}  
Here we obtain the expression for the distance between the spacecraft as  
function of tilt angle $\delta$. This can be computed in a straight forward way from   
Eq. (1) - (5) in the barycentric frame. The computations require less effort in the   
CW frame or in the $\{x', y', z'\}$  frame. In the $\{x', y', z'\}$ frame it is easy to   
see that the $z'$ contribution in the inter-spacecraft distance drops out since it is of   
higher order in $\a$. As remarked earlier we just need to consider $l_{12}$, the distance   
between SC 1 and SC 2, because of the symmetry assumed in the model. To the second order in   
$\a$ we get the following expression for $l_{12}$:  
\begin{equation}  
l_{12} (t, \d) = l + \Delta l_{12} (t, \d) \, ,  
\label{eq:l12}  
\end{equation}  
where,  
\be  
\Delta l_{12} (t, \d) = \frac{\a^2 R}{16 \sqrt{3}} \left[ 48 \left(\frac{3}{8} - \d_1 \right)   
- 15 \cos \theta + 48 \left(\frac{5}{8} - \d_1 \right) \cos 2 \theta - \cos 3 \theta \right] \, .  
\label{eq:dl12}  
\ee  
\begin{figure}[t]  
\centering  
\includegraphics[width=0.6\textwidth]{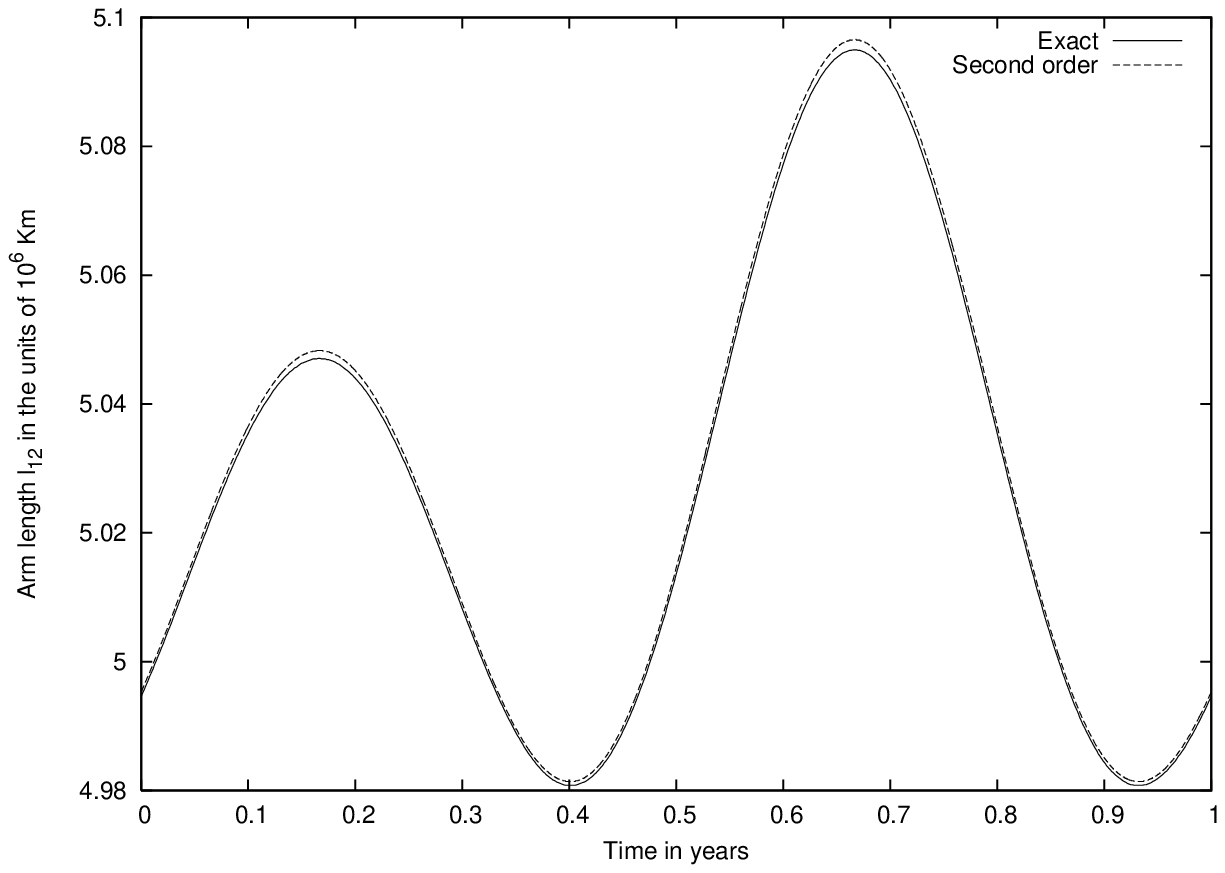}  
\caption{The variation of the lengths of the arms of LISA (the breathing modes) is shown   
in units of $10^6$ km.  The dashed curve shows the variation in $l_{12}$ over a year computed   
to the second order   
in $\alpha$, while the solid curve shows the exact variation. The discrepancy between the curves   
is less than 0.03 \%.}  
\label{brth}  
\end{figure}   
Here $\delta = \alpha\, \delta_1$ and $\theta = \O t - \pi / 3$.   
Since we restrict $\delta \sim o (\alpha)$, $\delta_1 \sim o (1)$.    
This expression is quite simple and easy to manipulate in order to extract the relevant   
information. Again, we see from figure \ref{maxvar} the exact orbits (solid curve) and the   
second order (dashed curve) approximation in $\a$ match very well. This fact motivates   
the analytic derivation of the equations of motion to second order in $\a$.   
We do this in section \ref{SC:3}. We make the following observations:  
\begin{itemize}  
\item To the first order in $\alpha$, $l_{12}$ is constant in time and equal to $l$.    
This result is known in earlier literature and was also obtained in paper I from the CW equations.   
It is only from the second order in $\a$ that the variation in arm-length appears.  
  
\item The variation in arm-length obtained up to the second order in $\alpha$ is close to   
the exact variation in arm-length. First we compare the variations in arm-lengths as a   
function of time for the  tilt angle of $\frac{\pi}{3}$ or when $\d = 0$. When $\d = 0$   
we get from Eq. (\ref{eq:dl12}),  
\be  
\Delta l_{12} (t, \d = 0) = \frac{\sqrt{3}}{32} \a l \left(6 - 5 \cos \theta +   
10 \cos 2 \theta - \frac{1}{3} \cos 3 \theta \right) \,.  
\label{eq:d=0}  
\ee  
With $\alpha = 1/60$ we have plotted $l_{12}(t, \d = 0)$, depicted by the dashed curve,   
in Fig. \ref{brth}. The dominant term is of double frequency - the term in $2\Omega t$   
- which gives two cycles in one year. The heights of the peaks are modulated by the   
single frequency term in $\Omega t$. Also the curve is `pushed' up by the constant term   
or the DC term. The term in $3 \O t$ is small and has little   
effect on the overall qualitative behaviour. The solid curve shows $l_{12} (t, \d = 0)$ for the exact   
Keplerian orbits. It is important to note that the second order result is very close to   
one obtained from the exact Keplerian orbits. The maximum difference in the flexing is   
about 0.03 percent.  
  
\item We plot the peak-to-peak amplitude of the flexing over the year as a function of   
$\delta$. This is shown in the Figure \ref{maxvar}.  
\begin{figure}[t]  
\centering  
\includegraphics[width=0.6\textwidth]{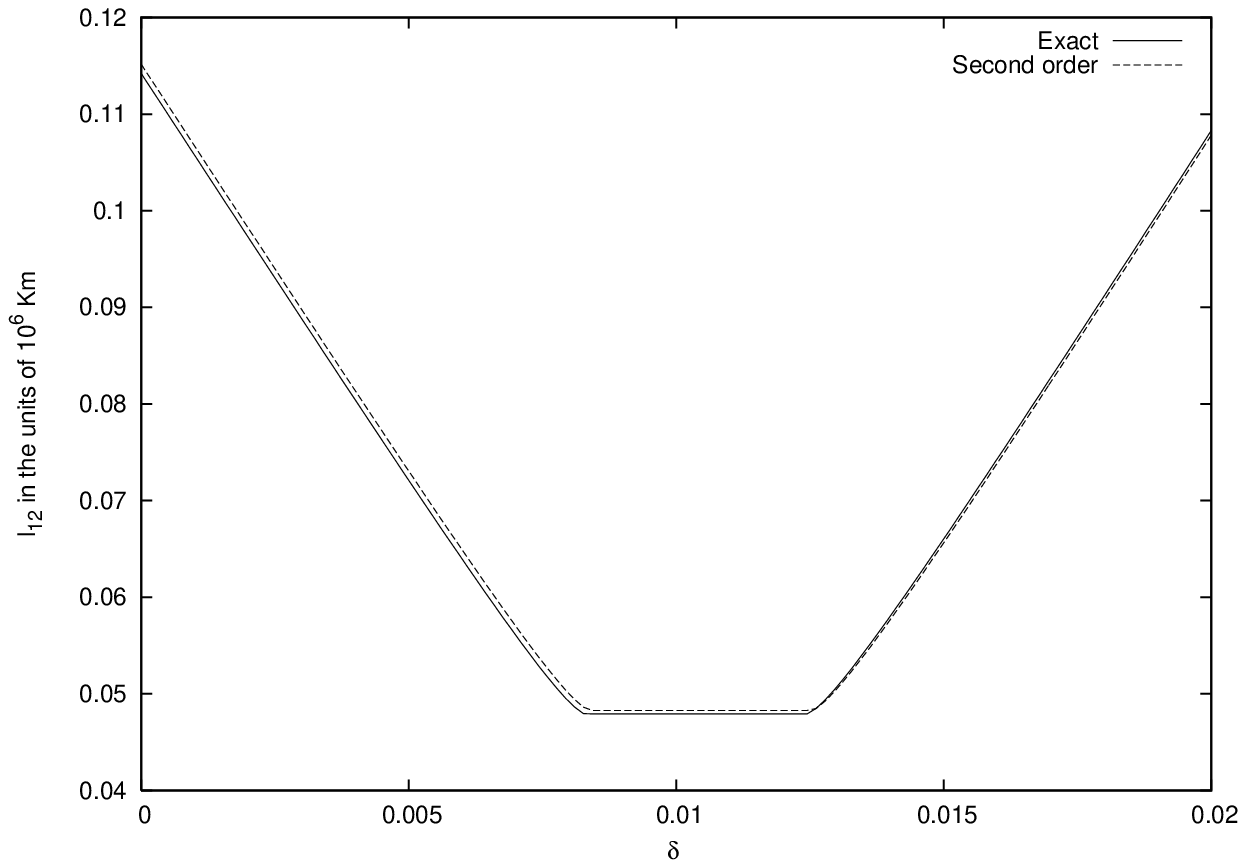}  
\caption{Maximum variation of arm length i.e. peak to peak amplitude of  
$l_{12}$ as function angle $\delta$, first order correction  around   
tilt angle $\frac{\pi}{3}$. }  
\label{maxvar}  
\end{figure}  
As is seen here, there is a range of values of $\delta$ for which the flexing is    
optimal, namely, $0.0083 \, < \,  \delta \, < \, 0.0125$. In this interval the curve appears    
flat; we show that the curve is flat in this region when computed to the second order in $\a$ and 
the exact model is numerically found to be similar. This can be inferred from Figure 2 (b) where we have plotted $l_{12}$ as a function of $t$ for various values of $\delta$, particularly when $\delta$ lies in the optimal range. From Figure \ref{maxvar}, it is observed that the difference between maximum and minimum $l_{12}$ remains constant $\sim 48000$ km, in the optimal range of $\d$. This result can be obtained exactly from Eq. (\ref{eq:dl12}) which is correct to second order in $\a$ as follows:
\par
Writing $x = \cos \theta$ and expanding the $\cos 2 \theta$ and $\cos 3 \theta$ in terms of 
$\cos \theta$, we obtain,
\be
\Delta l_{12} = - \frac{\a^2 R}{4 \sqrt{3}} \left[ x^3 - 24 \left (\frac{5}{8} - \d_1 \right) x^2 + 3 x + 3 \right] \,. 
\label{extrm}
\ee
The extrema of $\Delta l_{12}$ with respect to time can be obtained by setting the derivative of 
$\Delta l_{12}$ with respect to $x$ equal to zero. This yields an equation for $x$ parametrized by 
$\d_1$:
\be
x^2 - 2 (5 - 8 \d_1) x + 1 = 0 \,.
\ee
By evaluating the discriminant of this equation, we find that it has real roots only if $\d_1$ lies out of the interval $(0.5, 0.75)$. Within this interval 
the roots are complex and the extrema are obtained by setting $x = \pm 1$; it is clear from 
Eq. (\ref{extrm}) that the minimum of $l_{12}$ (or $\Delta l_{12}$) is attained when $x = 1$ and the maximum when $x = -1$. The peak-to-peak amplitude is just the difference between these two extrema and is easily computed from Eq. (\ref{extrm}). We find from these considerations that the peak to peak amplitude in the optimal region is:
\be
\Delta l_{12, {\rm max}} - \Delta l_{12, {\rm min}} = \frac {2 \a^2 R}{\sqrt{3}} = \frac {\a l}{\sqrt{3}} \sim 48,000 {\rm km} \,.
\ee
   
\item We also compute the r.m.s. variation in the arm-length from Eq. (\ref{eq:dl12})   
as follows:  
\be  
\langle \Delta l_{12} \rangle = \sqrt{3} \a^2 R \left( \frac{3}{8} - \d_1 \right) \, ,  
\ee  
where the bracket denotes the r.m.s. variation over 1 year. Clearly only the   
constant term contributes to the average. The variance of $l_{12}$ denoted by $V(l_{12})$   
is easily obtained by taking the sum of the squares of the coefficients of the time dependent   
terms because of the mutual orthogonality of the terms, and then dividing by 2.   
The division by the factor of 2 is because $\langle \cos^2 k \theta \rangle = 1/2,~ k = 1, 2, 3$.   
Thus,  
\be  
V(l_{12}(\d_1)) = \langle l_{12} - \langle l_{12} \rangle \rangle^2 (\d_1) =   
\frac{\a^4 R^2}{1536} \left[ 225 + 2304 \left( \frac{5}{8} - \d_1 \right)^2 + 1 \right] \, .  
\label{eq:var}  
\ee  
Clearly the variance is minimum when $\d_1 = 5 / 8$ which gives $\d = \a \d_1 = 0.0104$   
which is right in the centre of the optimal interval $[0.5, 0.75]$ of $\d_1$. 
From this value of $\d_1 = 5/8$, the minimum variance as well as the minimum r.m.s.   
amplitude of the breathing mode can be computed. The minimum r.m.s. amplitude of   
flexing is $\sim 16,000$ km. The gain factor $g$ in the r.m.s. amplitude with respect   
to the case $\d = 0$ is easily obtained from Eq. (\ref{eq:var}) to be,  
\be  
g = \sqrt \frac {V(l_{12}(\d_1 = 0))}{V(l_{12}(\d_1 = 5/8))} \sim 2.23 \, .  
\ee  
\end{itemize}  
With $\d_1=0$, the peak to peak amplitude is:   
\be  
\Delta l_{12,{\rm max}}- \Delta l_{12,{\rm min}} \ = \ l \times \frac{l}{R} \times   
\frac{\sqrt{3}}{2}\, \left[4\sqrt{6}-9\right] \ \sim \ 115,000 \ {\rm km} \, ,   
\ee  
whereas in the optimal range for $\d_1$ it is reduced to:  
\be  
\Delta l_{12,{\rm max}}- \Delta l_{12,{\rm min}} \ = \ l \times \frac{l}{R} \times   
\frac{1}{2\sqrt{3}} \ \sim \  48,000  \ {\rm km} \, ,  
\ee  
which is a factor of about 2.4. The quantities we try to minimize are however the   
Doppler shifts, i.e. the time derivatives of the arm-length variations. The   
time derivative of $\Delta l_{12}$ is:  
\be  
\frac{d}{dt}\Delta l_{12} \ = \ \frac{\a^2 R \Omega }{16\sqrt{3}}\,  
\left[15 \sin \theta -96\left(\frac{5}{8}-\delta_1 \right) \sin 2 \theta + 3\sin 3\theta   
\right] \,.  
\ee  
We can also compute the variance of $d/dt (\Delta l_{12})$:  
\be  
V\left[ \frac{d}{dt}\Delta l_{12}\right] \ = \  
\frac{\a^4 R^2 \Omega^2}{1536} \left[234+9216\left(\frac{5}{8}-\delta_1 \right)^2\right] \,.  
\ee  
It is obviously a minimum for $\delta_1=5/8$, but we note that the gain in  
Doppler shift is now:  
\be  
g_D \ = \ \sqrt{\frac{3834}{234}} \ \sim \ 4.05 \,,  
\ee  
so that it is even larger than the gain on the flexing itself. If we now investigate   
the peak to peak value of the Doppler shift, we get about 45.8 m/s for the   
standard ($\delta=0$) case, and only 8.2 m/s for the optimised case, so that the  
ratio is about 5.6, which seems interesting to note. The relative velocity is shown in Figure 
\ref{dplr} where its peak to peak value is apparent. It is seen from the figure that there are long periods of constant Doppler shifts for the optimised case. These results could be useful in the context of TDI.
\begin{figure}[t]  
\centering  
\includegraphics[width = 0.6 \textwidth]{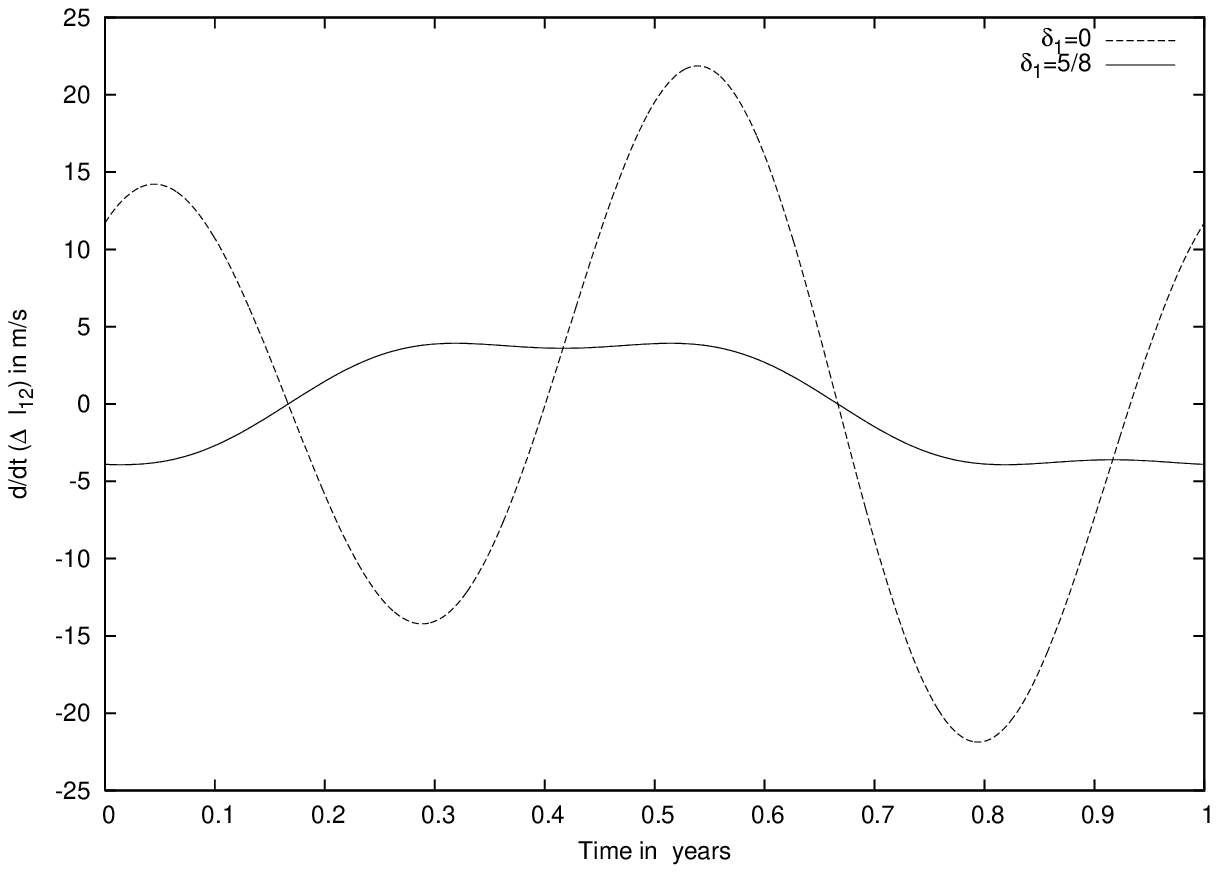}  
\caption{The relative velocity $\frac{d}{dt} (l_{12})$ is shown for the two cases $\d_1 = 0$ (dashed curve) and the optimised case $\d_1 = 5/8$ (solid curve) over the course of a year. For the optimised 
case the gain in peak to peak variation in relative velocity is $\sim 5.6$.}  
\label{dplr}  
\end{figure}  
\section{The Equations of Motion of Spacecraft to Octupole order \label{SC:3}}  
\subsection{Geodesic deviation equation to second order in the separation vector\label{SC:3A}}  
  
We begin with the Newtonian geodesic deviation equation to the first order in the separation vector:  
\be   
\ddot{\eta}^{a} + K^{a}_{b}\eta^{b} = 0,  
\label{gde1}  
\ee  
where $\eta^{a}(t)$, are the components of the $3$-vector separating the two particles   
at any time $t$ and $K^{a}_{b} = \Phi^{,a}_{,b}$. The $\Phi (r) = - k/r,~ k = G M$, is   
the  Newtonian gravitational potential, where $r$ is radial coordinate with the centre   
of force at the origin, $G$ the Newtonian gravitational constant and $M$ the mass of the Sun.   
Here the commas denote partial derivatives with respect to the coordinates. Upper and lower   
indices have been used to facilitate the summation convention.  
   
If the reference orbit is circular with radius $R$, the geodesic deviation equation in   
the barycentric frame is given by:   
\be  
\ddot{\eta}^{a} +\frac{ k}{R^5} (R^2 \delta^{a}_{b} -3 X^{a} X_{b}) \eta^{b} = 0,   
\ee  
where $X^a$ is the vector $(X, Y, Z)$. In vector notation (a bar over the quantity denotes   
a vector quantity and a hat denotes a unit vector),   
\be  
\ddot{\bar{\eta}} +\Omega ^{2}\bar{\eta}- 3 \Omega ^{2} \hat{R} (\hat{R}.\bar{\eta}) = 0,  
\label{qdrpl}  
\ee  
where $\Omega^2 = k/R^3$ and $\hat R$ denotes a unit vector in the radial direction.  
\par  
  
As seen before, the CW frame is a suitable frame. In this frame, the separation vector   
$\bar{\eta}$ is related to the CW coordinates by,  
\bea  
x & =& \eta^{x} \cos \Omega t + \eta^{y} \sin \Omega t \, , \no \\   
y & =& -\eta^{x} \sin \Omega t + \eta^{y}\cos\Omega t \, , \no \\  
z & =& \eta^{z},  
\label{CWT}  
\eea   
Then transforming Eq.(\ref{qdrpl}) to CW frame, it is found to be identical to the CW   
equations:   
\bea   
\ddot{x} - 2 \Omega \dot{y} -3 \Omega^2 x = 0 \, , \no\\  
\ddot{y} - 2 \Omega \dot{x} = 0 \, , \no\\  
\ddot{z} +  \Omega^2  z = 0.  
\label{CW}  
\eea  
However, in order to obtain the flexing of the arms we need to go to higher order in $\a$.   
We go to the second order in $\a$ or which is equivalent to the second order in separation   
vector $\bar \eta$. The second order equation contains octupolar terms of the gravitational   
field which essentially drive the flexing of the arms. The Taylor expansion of the Newtonian   
gravitational potential to the third order leads to the following equation for the separation   
vector $\bar \eta$:   
\be   
\ddot{\eta}^{a} + \Phi^{,a}_{,b}\eta^{b}+\frac{ 1}{2}  \Phi^{,a}_{,b c}\eta^{b}\eta^{c} = 0,  
\ee  
where   
\be \Phi^{,a}_{,b c} = - \frac{3 k}{R^5} (\delta^{a}_{b} X_{c} + \delta^{a}_{c} X_{b} +   
\delta_{bc} X^{a}) + \frac{15 k}{R^7}  X^{a} X_{b} X_{c}.  
\ee  
Or in the vector form, the above equation can be written as:  
\be  
\ddot{\bar{\eta}} + \Omega ^{2} \bar{\eta}- 3 \Omega^2\hat{R}  (\hat{R}.\bar{\eta})-   
\frac{3 \Omega^2}{R}(\hat{R}.\bar{\eta})\bar{\eta}- \frac{3}{2R}\Omega ^{2} \eta^2 \hat{R}   
+\frac{15}{2R}  \Omega^2\hat{R}  (\hat{R}.\bar{\eta})^2  = 0.  
\label{octpl}  
\ee  
The octupole terms in the equation are quadratic in the separation vector.   
These equations take on a simple form when transformed to the CW frame described   
earlier in Eq. (\ref{eq:CW}).  
\par  
In the CW frame, Eq. (\ref{octpl}) simplifies to:  
\bea  
\ddot{x}-2 \Omega \dot{y} - 3 \Omega^2 x + \frac{3 \Omega^2}{2 R}(2 x^2 - y^2 -z^2)& =& 0 \,  
 , \no \\  
\ddot{y} + 2 \Omega \dot{x} -\frac{3 \Omega^2}{R} x y& = &0 \, , \no\\  
\ddot{z} +  \Omega^2 z -\frac{3 \Omega^2}{R} x z& = &0.  
\label{gde2}  
\eea  
These equations can also be thought of as higher (second) order CW equations.  
\par  
In order to display the order of the various terms in $\a$, it is useful to employ   
dimensionless coordinates. We define the dimensionless variables in units of $l$ as,  
\be  
\tilde{x} = \frac{x}{l} \,, \tilde{y} = \frac{y}{ l} \,, \tilde{z} = \frac{z}{l} \,,   
\tilde{\rho} = \frac{\rho}{l},  
\ee  
where $\tilde{\rho}^2 = \tilde{x}^2+\tilde{y}^2+\tilde{z}^2$.   
\par  
We also set $\Omega$ to be unity for convenience. In the dimensionless variables,   
the equations (\ref{gde2}) reduce to:   
\bea  
\ddot{\tilde{x}} - 2 \dot{\tilde{y}} - 3\tilde{x} +3 \alpha (3 \tilde{x}^2-   
\tilde{\rho}^2)& =& 0 \, , \no \\  
\ddot{\tilde{y}} +2\dot{\tilde{x}} - 6 \alpha\tilde{x} \tilde{y}& =& 0 \, , \no \\  
\ddot{\tilde{z}} + \tilde{z} - 6 \alpha\tilde{x} \tilde{z}& =& 0 \, .  
\label{oct}  
\eea  
  
\subsection{The Equations of Motion via the Lagrangian \label{SC:3B}}  
  
The equations of motion can also derived with the Lagrangian for a central force   
expanded around a point on the reference orbit. The Lagrangian is:  
\begin{eqnarray}  
\mathcal{L} & = & \frac{1}{2}\left(\dot{X}^{2}+\dot{Y}^{2}+\dot{Y}^{2}\right)+  
\frac{k}{\left(X^{2}+Y^{2}+Z^{2}\right)^{1/2}} \, .  
\label{eq:kepL}  
\end{eqnarray}  
Here the mass of the test particle is ignored; it is not important  
for the central force problem. However if control terms and force  
terms are added then the mass of the test particle must be included. Here we ignore these   
effects. We carry out our analysis in the CW frame. The Lagrangian given by Eq.(\ref{eq:kepL})   
in the CW-frame takes the form:  
\be  
\mathcal{L} = \frac{1}{2}\left\{ \dot{x}^{2}+\dot{y}^{2}+\dot{z}^{2}+   
\left(R+x\right)^{2}+y^{2} \right\} +  \dot{y}\left(R+x\right)-  
\dot{x}y  + \frac{k}{\left[\left(R+x\right)^{2}+y^{2}+z^{2}\right]^{1/2}}\,.  
\label{eq:KLCW}  
\ee  
Then the force free Lagrangian equations of motion are given by:  
\begin{eqnarray}  
\ddot{x}-2 \dot{y}- (1 - \beta)(R+x) & = & 0\,, \no \\  
\ddot{y} + 2 \dot{x}- (1 - \beta) y & = & 0\,, \no \\  
\ddot{z}+ \beta z & = & 0\, ,  
\label{Lgr}  
\end{eqnarray}  
where,  
\be  
\beta = \left[1+\frac{2x}{R}+\left(\frac{x}{R}\right)^{2}+\left(\frac{y}{R}\right)^{2}+  
\left(\frac{z}{R}\right)^{2}\right]^{-\frac{3}{2}} \, .  
\ee  
Expanded to the octupole order, the quantity $\beta$ is given by:  
\begin{equation}  
\beta \simeq 1-\frac{3x}{R}+\frac{1}{2}\left[12\left(\frac{x}{R}\right)^{2}-  
3\left(\frac{y}{R}\right)^{2}-3\left(\frac{z}{R}\right)^{2}\right]\,.  
\label{eq:expand}  
\end{equation}  
To this order of approximation, Eq. (\ref{Lgr}) are identical to Eq. (\ref{gde2}).  
  
\section{The solutions to the Equations of Motion \label{SC:4}}  
  
We first obtain a general solution to the equations of motion in terms of arbitrary   
constants and then describe the family of solutions with tilt $\d \sim o(\a)$, which matches with the exact Keplerian orbits  Eq. (\ref{tltorb}) to second order in $\a$.   
Finally, we show that the variation in arm-length is optimal in two ways: (i) r.m.s sense and 
(ii) peak to peak sense. This is achieved by optimally choosing the arbitrary constants   
in the general solution. These results are valid in the neighbourhood of the exact orbits considered in section \ref{SC:2A}.  
 
\subsection{The General Perturbative Solution around $\delta = 0$ \label{SC:4A}}  
  
Accordingly, we write, $\tilde{x}  = \tilde{x}_0+\alpha  \tilde{x}_1$, $\tilde{y} =    
\tilde{y}_0 +\alpha \tilde{y}_1$, $\tilde{z} = \tilde{z}_0 +\alpha \tilde{z}_1$,  
 where,  $ \tilde{x}_0$, $\tilde{y}_0$ and $ \tilde{z}_0$ are solutions to the equations   
of motion to quadrupole order. These solutions satisfy certain criteria which we describe below.  
\par  
To the first order in $\alpha$ the $x_0, y_0, z_0$ satisfy the CW equations (\ref{CW}).   
As shown in paper I, the solutions of these equations satisfying no drift, no offset   
and which have the test-particle equidistant from the origin are the following:   
\bea  
\tilde{x}_0 &=&\frac{1}{2} \tilde{\rho}_0 \cos(t + \phi_0) \, , \no \\  
\tilde{y}_0 &=& -\tilde {\rho}_0 \sin(t + \phi_0) \, , \no \\  
\tilde{z}_0 &=& \frac{\sqrt{3}}{2}\tilde{\rho}_0 \cos(t + \phi_0) \, ,  
\label{cws}  
\eea  
where $\tilde {\rho}_0$ and $\phi_0$ are fixed for a specific orbit. For the LISA spacecraft,   
$\tilde{\rho}_0  = \frac{1}{\sqrt{3}}$. The orbits of the spacecraft are easily obtained by   
the appropriate choice of $\phi_0$. For convenience in solving the equations,   
we choose $\phi_0$ to be   
zero. The orbits with non-zero $\phi_0$ are easily obtained by translation.  
\par  
To the second order in $\alpha$, we get the equations for $\tilde{x}_1$,$\tilde{y}_1$  
 and $\tilde{z}_1$ as,   
\bea  
\ddot{\tilde{x}}_1 - 2 \dot{\tilde{y}}_1 - 3 \tilde{x}_1 + 3 (3 \tilde{x}_0^2 -  
 \tilde{\rho}_0^2) = 0 \, , \no \\  
\ddot{\tilde{y}}_1 + 2 \dot{\tilde{x}}_1 - 6 \tilde{x}_0 \tilde{y}_0 = 0 \, , \no \\   
\ddot{\tilde{z}}_1 + \tilde{z}_1 - 6 \tilde{x}_0 \tilde{z}_0 = 0.  
\label{prtb}  
\eea  
We first solve the $y$-equation. We substitute the solutions for $\tilde{x}_0$ and    
$\tilde{y}_0$ from the CW solutions (\ref{cws}), so that the $y$-equation becomes:  
\be  
\ddot{\tilde{y}}_1 + 2 \dot{\tilde{x}}_1 + \frac{3}{2} \tilde{\rho}_0^2 \sin 2t = 0 \, .  
\ee  
Integrating we have,  
\be  
\dot{\tilde{y}}_1 + 2\dot {\tilde{x}}_1 - \frac{3}{4} \tilde{\rho}_0^2 \cos 2 t = A \, , 
\label{y1dot}  
\ee  
where A is an integration constant. We substitute  $\dot{\tilde{y}}_1$ in the equation for  
 ${\tilde{x}_1}$ to obtain:  
\be  
\ddot{\tilde{x}}_1 + \tilde{x}_1 = \frac{1}{8} \cos^2 t  + 2 A + \frac{5}{8} \, .  
\ee  
Then integrating we get,  
\be  
\tilde {x}_1 =  B \cos t + C \sin t - \frac{1}{24} \cos2 t + 2 A + \frac{5}{8} \, .  
\ee  
Finally, integrating the equation for $\ddot{\tilde{z}}_1$ ,we get,  
\be  
\tilde {z}_1 =  E \cos t + F \sin t - \frac{1}{4 \sqrt{3}} \cos 2 t +  \frac{\sqrt{3}}{4}.  
\ee  
After integrating the ${\tilde y}_1$-equation, namely, Eq. (\ref{y1dot}) , we collect the solutions:  
\bea  
\tilde {x}_1 &=& B \cos t + C \sin t - \frac{1}{24} \cos 2 t + 2 A + \frac{5}{8} \, , \no \\  
\tilde {y}_1 &=& -2 B\sin t + 2 C \cos t + \frac{1}{6} \sin 2 t - 3 A t - \frac{5}{4} t +   
D \, , \no \\  
\tilde {z}_1 &=& E \cos t + F \sin t - \frac{1}{4 \sqrt{3}} \cos 2 t +  \frac{\sqrt{3}}{4}.  
\label{gns}  
\eea  
Here, $ A,B,C,D,E$ and $F$ are the integration constants and (\ref{gns}) is the general solution for  the perturbations to second order in $\a$. These constants can be determined from initial conditions. However, the term linear in $t$ in the solution for $\tilde {y}_1$ represents an unbounded drift. Since such solutions imply instability, the drift must be removed. This is achieved if we set   
$ A = -\frac{5}{12}$.   
  
\subsection{The family of solutions with varying tilt near $\d = 0$ \label{SC:4B}}  
  
In this subsection we show that the family of solutions with varying tilt $\d \sim o(\a)$ is consistent 
with the general solution obtained in Eq. (\ref{gns}). For this purpose we consider the exact orbits of the spacecraft given by Eq. (\ref{tltorb}) - (\ref{orbits}). We are therefore choosing a specific set of orbits for which the spacecraft distances are constant and equal to $l$ to the first order in the eccentricity (or equivalently $\a$). If these  orbits are expanded to second order in   
$\alpha$ (note $\delta = \a \d_1$ where $\d_1 \sim o(1)$), the following equations can be deduced for spacecraft 1 in the CW frame in which the general solution Eq. (\ref{gns}) was obtained:  
\begin{eqnarray}  
x & = & \frac{R \alpha}{\sqrt{3}}\,\cos t + 2 R \alpha^{2} \left[\left(\frac{1}{4} - \frac{\d_1}{2} \right) \cos t -\frac{1}{24}\cos 2t -\frac{5}{24} \right]\,,\nonumber \\  
y & = & -2 \frac{R \alpha}{\sqrt{3}}\,\sin t + 2 R \alpha^{2} \left[ \left( \d_1 -\frac{1}{2} \right)   \sin t +\frac{1}{6}\sin 2t \right]\,,\nonumber \\  
z & = & R \alpha\,\cos t + 2 R \alpha^{2} \left[\frac{1}{2\sqrt{3}} \left(\d_1 - 1 \right) \cos t -  
\frac{1}{4\sqrt{3}}\cos 2t + \frac{\sqrt{3}}{4} \right]\,.  
\label{eq:incwfr}  
\end{eqnarray}  
They match with Eq.(\ref{gns}) if the constants are chosen as follows:   
$ A=-\frac{5}{12},~ B=\frac{1}{4} - \frac{1}{2} \d_1,~ C=0, ~D=0, ~E=\frac{1}{2 \sqrt{3}} (\d_1 - 1)$ and $ F=0$.  
\par  
Constructing the full solution (adding first and second order solutions) from   
Eq. (\ref{cws}) and Eq. (\ref{gns}), we obtain identical solutions,  
\bea  
x_1 &=& \frac{l}{2 \sqrt{3}}\cos t +\frac{l^2}{2 R} \left[ \left(\frac{1}{4} - \frac{\d_1}{2} \right) \cos t - \frac{1}{24}\cos 2t -\frac{5}{24} \right] \, , \no \\  
y_1  &=& - \frac{l}{ \sqrt{3}}\sin t +\frac{l^2}{2 R} \left[ \left( \d_1 -\frac{1}{2} \right) \sin t +  
\frac{1}{6}\sin 2t \right] \, , \no \\  
z_1 &=& \frac{l}{2}\cos t + \frac{l^2}{2 R}\left[ \frac{1}{2\sqrt{3}} \left(\d_1 - 1 \right) \cos t -  
\frac{1}{4\sqrt{3}}\cos 2t + \frac{\sqrt{3}}{4} \right] \, ,   
\eea  
where we have now switched back to coordinates having dimensions of length and $x_1, y_1, z_1$   
being the coordinates of spacecraft 1. The coordinates of spacecraft 2 and 3 in the CW frame   
are obtained by replacing $t$ by $t-2 \pi/3$ and $t-4 \pi/3$ respectively.   
  
\subsection{Establishing the optimality of the solution \label{SC:4C}}  
  
We have shown that the second order CW equations are sufficient for an accurate description  
of the breathing of LISA. We are therefore allowed to use the general solution found  
in section \ref{SC:4A} for the purpose of optimisation. If we compute the distance   
between SC 1 and SC 2 using the general solution, with the same arbitrary constants for all   
SC, we find:  
\bea  
l_{12} &=& l \left[ 1+\frac{\alpha \sqrt{3}}{2} \left( \frac{1}{2}(5B+\sqrt{3}E) -   
\frac{5}{16} \cos\theta + \frac{1}{2}(3B-\sqrt{3}E)\cos 2\theta \right. \right. \no \\   
       && \left. \left. + \frac{1}{2}(3C-\sqrt{3}F)\sin 2\theta -\frac{\cos 3\theta }{48}  
 \right) \right] \, .
\label{dstgn}  
\eea  
The time varying part of $l_{12}$ is now:  
\be  
l_{12} - \langle l_{12} \rangle \ = \ \frac{l \alpha \sqrt{3}}{2} \left[  
- \frac{5}{16} \cos\theta + \frac{1}{2}(3B-\sqrt{3}E)\cos 2\theta  
+ \frac{1}{2}(3C-\sqrt{3}F)\sin 2\theta  
-\frac{\cos 3\theta }{48} \right] \,.
\label{lvar}  
\ee  
Since the terms are orthogonal, the minimisation of the variance amounts to minimising  
each coefficient separately.  The arbitrary constants, namely, $B, C, E, F$ occur only in   
the coefficients of the $\cos 2 \theta$ and $\sin 2 \theta$ terms while the rest of the terms   
have constant coefficients. Thus the minimisation amounts to setting the coefficients of   
$\cos 2 \theta$ and $\sin 2 \theta$ equal to zero. This gives the minimum r.m.s. variation of about 16,000 km which was the same result obtained at the end of section \ref{SC:2}.   
\par  
If we now introduce different constants  
B,C,E,F for each SC, it is easy to show that only terms in $\cos k \theta ,  ~~~ k=1,2,3$   
and $\sin 2\theta$ appear in the distance, in such a way that only the coefficients  
of $\cos 2\theta$ and $\sin 2 \theta $ can be made to vanish. The result is then exactly  
as the preceding one, showing that this is indeed a true r.m.s. optimum in the neighborhood of the exact solution that we have assumed. This is a general result. 
\par   
We also minimise the peak to peak flexing by considering the general solution. First, we consider the particular case when $3 C - \sqrt{3} F = 0$. Then as in section \ref{SC:2},
\be
\Delta l_{12} = - \frac{\a l \sqrt{3}}{24} \left[ x^3 - 12 (3B - \sqrt{3}E) x^2 + 3 x - 12 B \right] \,,
\label{vard}
\ee
where $x = \cos \theta$. Following the analysis analogous to section \ref{SC:2}  we set the derivative of 
$\Delta l_{12}$ with respect to $x$ equal to zero which yields the quadratic equation:
\be
x^2 - 8(3 B - \sqrt{3} E) x + 1 = 0 \, .
\ee   
This equation has complex roots when $ -0.25 < 3B - \sqrt{3} E < 0.25$ implying that the extrema of 
$ \Delta l_{12}$ are attained when $x = \pm 1$ and when this condition is satisfied. From Eq. (\ref{vard}) one obtains the minimum peak to peak variation as $\a l / \sqrt{3} \sim 48000$ km identical to the one found in section \ref{SC:2}. However, this family is larger, having two parameters $B$ and $E$ than the one considered with varying tilt which has just one parameter $\d$. In particular, we verify that for the family of solutions with varying tilt $\d$ investigated in section \ref{SC:4B}, the condition $ -0.25 \leq 3 B - \sqrt{3} E  \leq 0.25 $, after substituting their values in terms of $\d_1$, leads to, $0.5 \leq \d_1 \leq 0.75$ in agreement with the result obtained earlier in section \ref{SC:2}. 
\par
If we do not apriori assume that $3 C - \sqrt{3} F$ is zero, then the simple analysis as performed above, fails. We therefore resort to numerical methods. We first observe that the time varying part of 
$l_{12}$ up to the constant $l \a \sqrt{3} /4$ from Eq. (\ref{lvar}) is given by:
\be
F (a, b ; \theta) = a \cos 2 \theta + b \sin 2 \theta - \frac{5}{8} \cos \theta - \frac{1}{24} \cos 3 \theta \,,
\ee
where, $a = 3 B - \sqrt{3} E$ and $b = 3 C - \sqrt{3} F$. The peak to peak value of $F$ we denote by 
$\Delta F$. We observe that $F$ satisfies the following symmetries: 
$F (- a, b; \theta) = - F (a, b; \pi - \theta)$ and $F(a, -b; \theta) = F (a, b; - \theta)$ implying 
$\Delta F (a, -b) = \Delta F (-a, b) = \Delta F (a, b)$. Thus $\Delta F$ is an even function of $a, b$. 
We can therefore restrict ourselves to non-negative values of $a$ and $b$ when searching for the optimum 
value of $\Delta F$. Also for large $a, b$, $\Delta F \sim 2 \sqrt{a^2 + b^2}$, so that the minimum of 
$\Delta F$ must lie in the neighborhood of the origin with $a, b \sim o(1)$. A straight forward numerical computation shows that the minimum value of $\Delta F$ occurs when $ -0.25 \leq a \leq 0.25$ and $b = 0$, giving the minimum peak variation of $\sim 48,000$ km as before. This is again a general result.

\section{Concluding Remarks \label{SC:CR}}  
   
In this work we have studied the variation of the lengths of the arms of LISA in the   
neighbourhood of the exact solution that gives constant armlengths to first order in $\a$.   
We investigate the flexing of arms of LISA by going to the second order in the parameter $\a$.   
We numerically establish that the second order results provide a near accurate description   
of the orbits and the flexing of the arms. Because the second order expressions are tractable   
and amenable to analytic treatment, a lot of information can be gleaned from the expressions   
themselves. We show that by slightly changing the tilt angle of the plane of LISA from the   
proposed one of $60^\circ$ with the ecliptic, one can reduce the flexing of the arms.   
This is investigated with the aid of the exact Keplerian orbits as well as by analysing to second   
order in $\a$. We then obtain the tilt angle for which the  flexing of the arms is minimum. 
We find that the peak to peak amplitude of flexing is reduced by a factor of about 2.4. More 
importantly, in the context of application of TDI techniques, we show that the peak to peak Doppler 
shift is reduced by a factor of 5.6.
\par  
Secondly, we obtain the equations of motion to second order in $\a$ from the Newtonian geodesic   
deviation equation to the quadratic order in the separation vector, which involves  the   
octupolar expansion of the Newtonian gravitational potential of the Sun. These equations can   
also be obtained with a Lagrangian approach. The general solutions to these equations are found    
perturbatively, in the neighbourhood of the first order solution for which the armlengths are   
constant to first order in $\a$. We then check that the family of solutions with varying tilt is   
reproduced from the general solution by choosing the arbitrary constants appropriately. Finally 
we use the general solution to establish that our solution of the minimum flexing of the arms is more generally valid in the neighborhood of the exact solution assumed. This is important from the point of view of a realistic launch because the flexibility in the choice of parameters allows for some error, which is inevitable, and this is possible without sacrificing performance.
      
\begin{acknowledgments}
SK would like to acknowledge DST, India for the WOS-A.     
\end{acknowledgments}
\vspace{24pt}   
%
%
%
%
  
\end{document}